\documentstyle[11pt]{article}
\def\hybrid{\topmargin 0pt      \oddsidemargin 0pt
	\headheight 0pt \headsep 0pt
	\textheight 9in         
	\textwidth 6.25in       
	\marginparwidth .875in
	\parskip 5pt plus 1pt   \jot = 1.5ex}

\catcode`\@=11
\def\marginnote#1{}
\newcount\hour
\newcount\minute
\newtoks\amorpm
\hour=\time\divide\hour by60
\minute=\time{\multiply\hour by60 \global\advance\minute by-\hour}
\edef\standardtime{{\ifnum\hour<12 \global\amorpm={am}%
	\else\global\amorpm={pm}\advance\hour by-12 \fi
	\ifnum\hour=0 \hour=12 \fi
	\number\hour:\ifnum\minute<10 0\fi\number\minute\the\amorpm}}
\edef\militarytime{\number\hour:\ifnum\minute<10 0\fi\number\minute}

\def\draftlabel#1{{\@bsphack\if@filesw {\let\thepage\relax
   \xdef\@gtempa{\write\@auxout{\string
      \newlabel{#1}{{\@currentlabel}{\thepage}}}}}\@gtempa
   \if@nobreak \ifvmode\nobreak\fi\fi\fi\@esphack}
	\gdef\@eqnlabel{#1}}
\def\@eqnlabel{}
\def\@vacuum{}
\def\draftmarginnote#1{\marginpar{\raggedright\scriptsize\tt#1}}

\def\draft{\oddsidemargin -.5truein
	\def\@oddfoot{\sl preliminary draft \hfil
	\rm\thepage\hfil\sl\today\quad\militarytime}
	\let\@evenfoot\@oddfoot \overfullrule 3pt
	\let\label=\draftlabel
	\let\marginnote=\draftmarginnote
   \def\@eqnnum{(\theequation)\rlap{\kern\marginparsep\tt\@eqnlabel}%
\global\let\@eqnlabel\@vacuum}  }

\def\numberbysection{\@addtoreset{equation}{section}
	\def\theequation{\thesection.\arabic{equation}}}

\def\underline#1{\relax\ifmmode\@@underline#1\else
	$\@@underline{\hbox{#1}}$\relax\fi}

\def\titlepage{\@restonecolfalse\if@twocolumn\@restonecoltrue\onecolumn
     \else \newpage \fi \thispagestyle{empty}\c@page\z@
	\def\thefootnote{\fnsymbol{footnote}} }

\def\endtitlepage{\if@restonecol\twocolumn \else  \fi
	\def\thefootnote{\arabic{footnote}}
	\setcounter{footnote}{0}}  
\catcode`@=12
\relax

\def\beq{\begin{equation}}
\def\eeq{\end{equation}}
\def\bea{\begin{eqnarray}}
\def\eea{\end{eqnarray}}
\def\bar{\overline}
\def\z{{\bar {z}}}
\def\nn{\nonumber}
\def\pa{\partial}
\def\d{{\cal D}}

\def\N{ {\Psi}}
\def\P{F}
\def\Pb{{\bar{F}}}

\def\m{\mu}
\def\n{\nu}
\def\rh{\rho}
\def\s{\sigma}

\def\th{\theta}
\def\m{\mu}
\def\e{\epsilon}

\def\a{\alpha}
\def\b{\beta}
\def\g{\gamma}
\def\d{\delta}
\def\l{\lambda}
\def\demi{{1\over 2}}

\def\o{\omega}
\def\r{\rho}

\def\bel{\m^z_\z }
\def\belb{\m^\z_z }
\def\cz{c^z  }
\def\czb{c^\z  }
 \def\bzz{b_{zz} }
 \def\bbzz{b_{\z\z} }
\def\gr{\a^{\demi,0}_\z }
\def\grb{\a^{0,\demi }_z }
\def\gg{\g^{\demi,0}  }
\def\ggb{\g^{0,\demi }  }
 \def\gz{\b_{{3\over 2},0} }
 \def\gzb{\b_{  0, {3\over 2} } }
\def\pz{\pa_z}
\def\pzb{\pa_\z }

\def\ee{\eea}
\def\be{\bea}

\relax
\numberbysection
\hybrid
\begin{document}
\begin{titlepage}
\begin{center}
 \hfill   DAMTP/96-50, PAR--LPTHE 96--15 \\
 [.5in]
{\large\bf A UNIFYING TOPOLOGICAL  ACTION  FOR \\   HETEROTIC AND TYPE
II  SUPERSTRING  THEORIES}\\[.5in]
 {\bf  Laurent Baulieu  }\footnote{email address: baulieu@lpthe.jussieu.fr} \\
    	   {\it LPTHE\/}\\
       \it  Universit\'e Pierre et Marie Curie - PARIS VI\\
       \it Universit\'e Denis Diderot - Paris VII\\
Laboratoire associ\'e No. 280 au CNRS
 \footnote{ Boite 126, Tour 16, 1$^{\it er}$ \'etage,
        4 place Jussieu,
        F-75252 Paris CEDEX 05, FRANCE. }

        {\bf   Michael B.  Green }\footnote{email address:
M.B.Green@damtp.cam.ac.uk} \\
    	   {\it DAMTP, Cambridge\/}
 \footnote{ Silver Street,  Cambridge CB3 9EW, UK. }

{\bf Eliezer Rabinovici  }\footnote{email address:ELIEZER@vms.huji.ac.il} \\
    	   {\it Racah Institute of Physics \/}\\
       \it  Hebrew University, Jerusalem

\end{center}

\begin{quotation}
\noindent{\bf Abstract }    The heterotic and type II superstring actions are
identified  in different
anomaly-free decompositions of a single topological sigma-model action
depending on
bosonic and fermionic
coordinates, $X^\mu $ and $  \r^A $ respectively, and of their topological
ghosts. This   model
results from   gauge-fixing the  topological gauge symmetry  $\delta X^\mu =
\epsilon^\mu (z,\bar
z)$ ($\mu =1,2,\dots, 10$) and $\delta \r^\alpha= \epsilon^\alpha (z,\bar z)$.
($\alpha=1,2.\dots, 16$).    From
another viewpoint the heterotic and type II superstring actions  emerge as two
different gauge-fixings of  the same  closed   two-form.
Comments are also made concerning the possibility of relating  $\rho^\alpha$ to
a  Majorana-Weyl space-time spinor superpartner of $X^\mu$.

  \end{quotation}
 \end{titlepage}
 \newpage

%

\def\nn{\nonumber}

\def\nn{\nonumber}
\newpage\null
 \section{Introduction}

There have been various suggestions of an underlying topological basis for
string and superstring theory \cite{wittensigma}-\cite {Warner}.
In particular, the relevance of twisted nonlinear sigma models to mirror
symmetry
of Calabi--Yau spaces  and of the complexification of
space-time has   been  pointed out in
\cite{wittensigma} and  \cite{Berkovits}.
In this paper  we will pursue the idea of writing superstring actions  in the
\lq Neveu--Schwarz--Ramond' (NSR) formalism as  sums of exact and $s$-exact
terms
where $s$ denotes the BRST
transformation associated with local redefinitions of world-sheet fields,
$X^\mu(z,\bar z)$ (where $\mu = 1, \dots, 10$)  together with anticommuting
coordinates
$\rho^\alpha$ ($\alpha =1,2,\cdots,16$).  Thus, we will consider theories based
on the very large symmetry,
\be \label{fond}
\d X^\mu(z,\z)=\e^\mu_X(z,\z)
\nn\\
\d \rh^\a(z,\z)=\e^\a_\rh(z,\z).
  \ee

A classical action with such a huge symmetry is guaranteed to be purely
topological; any lagrangian density invariant under (\ref{fond}) must be
locally
a pure derivative.  The rest of the gauge-fixed quantum action is then
ghost-dependent and exact under the BRST transformation, $s$,    associated
with
the
symmetry
defined in (\ref{fond}). Once a suitable gauge is chosen the fields
$X^\mu$ will
describe the target-space coordinates and $\rho^\alpha$ will become the
internal
symmetry coordinates of the heterotic string.  The world-sheet fermionic
coordinates
associated with $(0,1)$ supersymmetry for the heterotic theory or $(1,1)$
supersymmetry for the type II theories arise as combinations of  ghosts and
antighosts
for this symmetry, as will be seen in section 2.  The symmetry (\ref{fond})
refers
to a given world-sheet.  In order to define the superstring theory one must
also
perform the usual sum over all super world-sheets that reduces to integration
over
the moduli space of two-dimensional metrics and gravitini.  These contribute
terms
in the string quantum action that are also guaranteed to be $s$-exact.

 Gauge-fixing will be discussed in section 3.  There we show that
the choice of superconformal gauge, together with  suitable choices  of gauge
functions
for $\rho^\alpha  $ and $X^\mu$ leads to an action, $I_B$, that is the sum of a
term that is
$s$-exact and a topological term.  This anomaly-free action  may be interpreted
as
the action for the heterotic string, $I_{het}$, by  writing it as $I_B =
I_{het} + I'$,
where  $I'$ involves fields that may be integrated out.  Alternatively, it may
be interpreted as the type II action, $I_{II}$, by writing   $I_B = I_{II} +
I^{\prime\prime}$, where  $I^{\prime\prime}$ also involves fields that may be
integrated out.  These decompositions  are consistent because the terms
$I_{het}$, $I_{II}$, $I'$ and $I^{\prime\prime}$ are separately free of
anomalies.  It is non-trivial that such decompositions  exist.   The
heterotic fermions are identified with  elements of the topological BRST
quartet of fields in the $\rh$ sector.

The fact  that  both the type II and heterotic superstring
actions can be viewed as  anomaly-free  truncations of the same  topological
$\sigma$
model with bosonic and fermionic coordinates in the superconformal gauge
means
that the full theory  contains purely topological observables. This topological
system  also determines
matter-dependent   observables characterized by the  cohomology of the
BRST symmetry   of the superconformal symmetry, which now appears as a small
part of the topological $\sigma$-model BRST symmetry.   This unification of the
type II and heterotic superstring actions requires the introduction of new
field variables which  decouple  from physical
quantities  but which play a r\^ole in the computation of the  purely
topological observables.

 The intertwining of world-sheet and space-time symmetries of superstring
theory
is an interesting and subtle issue.   It is tempting to believe that the  NSR
and the \lq Green--Schwarz'  (GS)
 formalism (which has manifest space-time supersymmetry) may be viewed as
different
gauge-fixed versions of a theory that somehow encompasses both.
  In that case there  could
be two anticommuting coordinates like $\rho^\alpha$, one of which has been
chosen to be zero in (\ref{fond}).  After a suitable gauge choice one or both
of these
coordinates may then
be identified with Majorana--Weyl space-time spinor
superpartners of $X^\mu$ in the heterotic or type II theories.   A version of
the BRST algebra that manifests this space-time supersymmetry will be presented
 in section 4 although we have not succeeded in obtaining the  GS action by
gauge
fixing in this manner.

 \section{The BRST algebra for world-sheet matter and supergravity}

The BRST symmetry corresponding to the gauge symmetry (\ref{fond}) is obtained
by changing the parameters $\e_X$ and $\e_\rh$ into
topological ghosts
$\P _X(z,\z)$ and $\P _\rh(z,\z)$  and by introducing the antighosts
$\Pb _X(z,\z)$ and $\Pb _\rh(z,\z)$ with their Lagrange multipliers $\l
_X(z,\z)$ and
$\l _\rh(z,\z)$.   The fields $\P_X$, $\Pb_X$ and $\l_\rh$ are fermionic while
$\P_\rh$, $\Pb_\rh$ and $\l_ X$ are bosonic fields. The graded differential
BRST operator
$s$  which encodes the topological  gauge symmetry (\ref{fond})  is defined as
\be sX^\m  &=& i\P^\m _X ,
\nn\\ s\P^\m _X &=&0,
\nn\\ s\Pb_X^\m  &=& \l_X^\m  ,
\quad\quad s\l^\m _X= 0,
\ee  and
\be s\rh^\a  &=& i\P^\a_\rh,
\nn\\ s\P^\a_\rh &=& 0,
\nn\\ s\Pb^\a _\rh &=& \l^\a _\rh,
\quad\quad  s\l^\a _\rh=0 \label{symmone}.
\ee
This symmetry is   needed to define a BRST invariant action
associated with the symmetry (\ref{fond}) on  a given worldsheet.  The
possibility of relating $\rho^\alpha$ to a space-time spinor coordinate will be
described in section 4.

The integration over the world-sheet metric and gravitino may be carried out in
a
superconformal gauge that fixes the super-Weyl and super-reparametrization
symmetries.  The result  is a theory that possesses the
BRST symmetry associated with  an $N=1$ superconformal theory .   In the
following we shall consider the situation in which there is no
conformal anomaly.  In that case the  conformal factors can be
gauged away  so that  the  variables that enter the gravitational part of the
action are the Beltrami differential
$\bel$ and its anticommuting reparametrization ghost
$\cz$, the conformally invariant part of the gravitino $\gr$ and its commuting
supersymmetry ghost $\gg$, together with the reparametrization and
supersymmetry
antighosts $\bzz$ and $\gz$ for the holomorphic sector and the complex
 conjugates for the other sector.  These fields  collectively constitute the
(super) Beltrami variables.  The use of these variables allows for a complete
separation between the left-moving and right-moving sectors.   The holomorphic
sector possesses the  factorized BRST symmetry algebra,
\be s\bel&=&\pzb\cz+\cz\pz\bel-\bel\pz\cz + 2i \gr\gg,
\nn\\s\gr&=&\pzb\gg+\demi \gg\pz\bel- \bel \pz\gg +\cz\pz\gr-\demi\gr\pz\cz,
\nn\\ s\cz&=&\cz\pz\cz+{i \over 2} \gg\gg,
\nn\\ s\gg&=&\cz\pz\gg + \demi\gg\pz\cz,\label{symmtwo}
\ee
with analogous equations in the anti-holomorphic sector.
The  superconformal gauge conditions are
    $\bel=\gr=0$ in the holomorphic sector and $\belb=\grb=0$ in the
antiholomorphic sector. It is well-known that in this gauge the ghost action is
given by
the $s-$exact term \cite{bellon},
\be
\label{bel} s\left(\bzz\bel+\gz\gr+\bbzz\belb+\gzb\grb\right) =-\bzz\pzb\cz+\gz
\pzb\gg -\bbzz\pz \czb+\gzb \pz \ggb.
\ee
 From now on we will work in  this gauge.

The issue of the consistency of this construction on world-sheets
of higher genus is not addressed here.

\section{ The $s$-exact and $d$-exact  action.}

We begin by considering the type II superstring action in a flat target-space
metric  expressed in the superconformal gauge,
\be \label{dec}  I_{II} =
\int d^2z\ \sum_{\m=1}^{10}
\left (
\pz X^\m\pzb X^\m -i\N^{\m1}\pzb \N^{\m1} -i\N^{\m2}\pz  \N^{\m2}
\right) + I_{II}^{ghost},\ee
where $I_{II}^{ghost}$ is the standard $(1,1)$ superconformal ghost action
(\ref{bel})  for the type II theories.

The field $X^\mu$ has conformal weight $(0,0)$ while the  conformal weight of
the components of the two-dimensional  world-sheet Majorana spinor, $\N^{\m
i}$, are  $(\demi,0)$ for
$i=1$ and  $(0, \demi) $ for
$i=2$. In this and subsequent formulae the signature of space-time will be
arbitrary.

The next step is to define new fermionic fields, $\P_X$ and
$\Pb_X$,  that are linear combinations of the NSR fields
by
\be
2\N^{a,1} &=& \Pb^{11-a}_X+\P_X^{ a}-i\P^{11-a}_X\nn\\
2i\N^{11-a,1} &=& \Pb^{11-a}_X-\P_X^{ a}+i\P^{11-a}_X
\ee
\be
2\N^{a,2} &=& \Pb _X^{a}+\P_X^{ a}+i\P^{11-a}_X\nn\\
2i\N^{11-a,2} &=& \Pb^{ a}_X-\P_X^{ a}-i\P^{11-a}_X\label{redef}
\ee
for $1\leq a\leq 5 $.
The notation indicates that the fields $\P_X$ and
$\Pb_X$ will be identified with  the topological  ghosts and antighosts
introduced in the last section.   These field redefinitions can be considered
to be  twists of the
original fields \cite{vafaone,wittensigma}.
The inverse relations
expressing  the topological ghosts and antighosts in  terms of the NSR fields
are
\be 2\P_X^{a} &=&  \N^{a,2}+\N^{a,1}-i\N^{11-a,1}-i\N^{11-a,2} \nn\\
2i \P_X^{11-a} &=&  \N^{a,2}-\N^{a,1} + i\N^{11-a,1} - i\N^{11-a,2} ,
\label{threefour}\ee
\be
 \Pb_X^{a} &=&   \N^{a,2}+ i\N^{11-a,2}  \nn\\
 \Pb_X^{11-a} &=&  \N^{a,1}+i \N^{11-a,1} .
\label{threefive}\ee
In the twisted version of the theory the  fields defined in (\ref{threefour})
have zero world-sheet spin  while the conjugate fields defined in
(\ref{threefive}) have spin one.\footnote{The apparent mismatch in the
conformal weights on the left-hand and right-hand sides of these equations can
be compensated by field redefinitions involving factors of  $\gg$ or
$\ggb$  \cite{twist}, as will be reviewed later.}

The identification of  $\P_X$ and
$\Pb_X$ with the topological ghosts and antighosts follows from the expression
for  $I_{II}$ in terms of the new fields,
\be\label{NSRR} I_{II}  & =&  \int d^2z \sum_{a=1}^5
  (\ -\l_X^a  \l_X^{11-a} +\l_X^a(\pz X^a+i\pz X^{11-a} ) +\l_X^{11-a}(\pzb
X^a-i\pzb X^{11-a} )\nn\\
&& -i\Pb_X^a (\pz \P^a+i\pz
\P^{11-a} ) -i\Pb_X^{11-a} (\pzb \P^a-i\pzb \P^{11-a} )) + I_{top} +
I_{II}^{ghost},
\ee
where
\be\label{top}
I_{top}=i\int \sum_{a=1}^5 dX^a\wedge dX^{11-a} = i \int
\sum_{a=1}^5
d( \ X^a\wedge dX^{11-a}).
\ee
This has a form that is manifestly the sum of a topological term,  $I_{top}$,
and a $s$-exact term,
\be\label{II}    I_{II}  &=&
\int d\ (  i\sum_{a=1}^5   X^a\wedge dX^{11-a}\ )   +\int d^2z\ s\left[
\bzz\bel+\bbzz\belb+\gz\gr+\gzb\grb  \right.
\nn\\
&&\left.  +   \sum_{a=1}^5 \left(
 \Pb_X^a(\demi  \l_X^{11-a} +\pz X^a+i\pz X^{11-a} )
+\Pb_X^{11-a}(\demi\l_X^a+\pzb  X^a-i\pzb X^{11-a} )\right)\right].
\ee
The
identification of the NSR fermions as combinations of the topological ghosts
and
antighosts of the gauge symmetry (\ref{fond}) is   quite striking.

 Although ghost number is not conserved in these definitions, it is conserved
modulo 2, which is all that is required in the quantum theory.
The fields $\P_X$ and $\Pb_X$  are assumed to have boundary values such  that
$\int d^2z \pz (\Pb\P)$ and $\int d^2z \pzb (\Pb\P)$
vanish (for instance with periodic or anti-periodic conditions).

We now turn to consider how the heterotic string action can be obtained by
gauge fixing the same topological symmetry.  In this case we will find that the
heterotic action can be identified with an anomaly-free part of a larger
action, $I_B$, that contains other fields that decouple from the fields in the
heterotic theory.  The starting point is the action for the heterotic string in
the superconformal gauge,
\be
\label{het}
I_{het} &=&
\int
d^2z\
\left (  \sum_{\m=1}^{10}
\pz X^\m\pzb X^\m
-i\sum_{\m=1}^{10} \N^{\m1}\pzb \N^{\m1}- \right. \nn\\
&& \quad \left. \bzz\pzb\cz+\gz \pzb\gg
  -i\sum_{i=1}^{32}f^{ i}\pz  f^{i}
-\bbzz\pz \czb\right) ,
\ee
where  the $(0,1)$ superconformal ghost action has been explicitly included.
The holomorphic sector   is the same as in  the type  II case while the
anti-holomorphic sector contains the 32 anticommuting
Majorana-Weyl world-sheet  spinors,
$f^{ i}$, in addition to the bosonic coordinates and the anti-holomorphic
$(b,c)$ ghost system.

In order to obtain this action from an expression that is a sum of $d$-exact
and $s$-exact pieces we next observe that the term   involving the fermionic
fields, $f^i$,  can be rewritten as
\be \label{hete}
\int d^2z\sum_{i=1}^{32}f^{ i}\pz  f^{i} =\int
d^2z\sum_{\a=1}^{16}\l_\rh^{\a}\pz  \rh^\a,
\ee where
\be\label{reredef}
f^\a={{\rh^\a +\l^\a}\over2} \nn\\
f^{33-\a }={{\rh^\a -\l^\a}\over{2i}}
\ee
for $1\leq \a\leq 16$. If  periodic or antiperiodic boundary spin structures
are chosen for  all $f^{ i}$  (which defines the $SO(32)$ heterotic string) the
  term
$\int d^2z\pz(f^i f^{33-i})$ vanishes. Otherwise (for example, in the
$E_8\times E_8$ case) there is an additional $d$-exact term on the right-hand
side of (\ref{hete}).

The expression (\ref{hete}) can be identified with part of the $s$-exact
action,
\be
I^{\prime\prime} &=&
i\int d^2 z\sum_{\a=1}^{16} s(\Pb_\rh^\a \pz\rh^\a)\nn\\
&=&
\int d^2 z\sum_{\a=1}^{16}  (
i\l_\rh^\a \pz\rh_\rh^\a-\Pb_\rh^\a \pz\P_\rh^\a).\label{primedact}
\label{newprime}\ee
This expression has no net propagating fields -- the bosonic ghosts and
antighosts ($\Pb_\rh^\a$ and $\P_\rh^\a$) balance the heterotic fermions
($\rho^\alpha$ and $\lambda^\alpha$).

In fact, since $I^{\prime\prime}$ is $s$-exact, the  action
\be\label{IH}
I_B  =  I_{II} + I^{\prime\prime},
\ee
 would have been an equally good action for the type II theories.  The fields
in $I^{\prime\prime}$ (the $\rh$ sector) simply decouple in the functional
integral for any correlation function of type II fields (which are in the $X$
sector).     For this to be consistent is essential that the terms $I_{II}$ and
$I^{\prime\prime}$ are separately free of anomalies, which is manifestly the
case (taking into account our
choice of equal   conformal weights for $\rh$ and $\P_\rh$).
  The construction is reminiscent of  the definition of   topological
Yang-Mills theory as
the
 BRST invariant gauge-fixing of  the second Chern class \cite{bsym}.

Less obvious is the fact that the action $I_B$ can also be broken up in another
anomaly-free manner,
\be I_B=I_{het}+I',
\label{decomtwo}\ee
where $I_{het}$  was defined in  (\ref{het}) and
\be  I'=\int d^2z
\left ( -i\sum_{\m=1}^{10} \N^{\m2}\pz  \N^{\m2}
 +\gzb \pz \ggb
  -\sum_{\a=1}^{16}\Pb_\rh^{\a}\pz  \P_\rh^{\a} \right).
\ee
The action $I'$  only involves anti-holomorphic fields that are absent from the
usual heterotic action.

$I_{het}$ and $I'$ are not separately BRST exact -- only their sum is. The
separation  of
$I_B$ into these two actions is however consistent and provides two independent
theories  because each of them is  free of gravitational and conformal
anomalies.  For $I_{het}$ this follows by the usual arguments.   That  $I'$  is
  independently  anomaly-free  follows  if we attribute
conformal weight $(0,\demi)$ to all the
fields  of the $\rh$ sector.  In that case the system
($\Pb_\rh^{\a},\P_\rh^{\a}$)  contributes  $-16$ to the conformal anomaly, the
NSR
fields  $\N^{\m2}$  contribute  $5$ and the
 ($\gzb, \ggb$) system contributes  $ 11$, giving a total conformal anomaly  of
$5+11-16=0$.\footnote{Recall that a system of conformal fields $(A,B)$ with
Lagrangian
$A\partial B$ has a conformal anomaly equal to $ \pm 2(6n^2-6n+1)$ where $n$ is
the
conformal weight of the field $A$ and the sign $+$ (  $-$) occurs if $A$ and
$B $ commute (anticommute) \cite{friedan}.}

The fields of the usual heterotic theory may be denoted by $ \phi_1 =\{X$,
$\N^1$, $\rh$, $\l_\rh$,  $\bzz$,  $\cz$, $\bbzz$, $\czb$,
$\gz$,
  $\gg\}$  and the fields in $I'$ by
$\phi_2=  \{ \N^2$, $\P_\rh$, $\Pb_\rh$, $\gzb$,   $\ggb\}$.    A heterotic
theory observable
$A(\phi_1)$ is defined by the functional
integral,
\be
\label{fact} \langle A(\phi_1) \rangle &=&\int [d\phi_1][d\phi_2] A(\phi_1)
\exp
i(I_{het}[\phi_1]+I_2[\phi_2])\nn\\  &=&{\it N}
\int [d\phi_1]  A(\phi_1) \exp i I_{het}[\phi_1],
\ee
where   ${\it N}$ is the partition function for the theory generated by $I'$.
It is an
irrelevant normalisation factor in the context of the usual  heterotic theory.
Equation  (\ref{fact})   relies on the fact that both the theories defined by
$I_{het}$ and $I'$ are
separately anomaly-free and are therefore truly decoupled.   It is  interesting
that there is a mapping  between the two theories  due to the existence of the
topological BRST symmetry which mixes their fields.

We have thus shown  that, up to
irrelevant terms,  the heterotic theory can be obtained from a $s$-exact action
with the same topological BRST symmetry as in the NSR formulation of the type
II theory.   The   difference between these physically different theories
arises from different anomaly-free eliminations of   fields.   This might be of
relevance in the context of the apparently rich set of interrelationships
between type II  and  heterotic theories.

The observables of the type II or heterotic models  are defined
by the residual symmetries that survive our choices of     gauge functions,
acting
on the remaining (twisted) fields rather than the full topological BRST
symmetry. However,   there are observables of the theory defined by the large
action, $I_B$, that  possess
the full topological BRST symmetry  of the type  encountered in
topological $\s$  models.

It is worth noting that the BRST algebra of the beltrami variables
(\ref{symmtwo}) can be interpreted as a   BRST algebra of topological 2-D
gravity by the explicit  change of variables,
\begin{equation}\label{redefs}
\Psi^z_{\bar z}=2i\gr\gg, \qquad \Phi^z={{i}\over{2}}\gg\gg,
\end{equation}
 in the holomorphic sector and corresponding definitions in the
anti-holomorphic sector.  The field $\Psi^z_{\bar z}$ is interpreted as  the
topological ghost of the Beltrami differential, while $\Phi^z$ is the ghost of
this ghost. The explicitly topological  BRST algebra on the Beltrami fields is
then,
 \be s\bel&=&\Psi^z_{\bar z}+\pzb\cz+\cz\pz\bel-\bel\pz\cz ,
\nn\\s\Psi^z_{\bar z}&=&\pzb\Phi^z+\Phi^z\pz\bel-\bel
\pz\Phi^z+\cz\pz\Psi^z_{\bar z}-\Psi^z_{\bar z}\pz\cz,
\nn\\ s\cz&=&\Phi^z+\cz\pz\cz ,
\nn\\ s\Phi^z&=&\cz\pz\Phi^z-\Phi^z\pz\cz \label{symmtop},
\ee
with corresponding equations in the anti-holomorphic sector.
One can thus view the ghost system  as originating from  gauge-fixing
topological 2-D gravity in the gauge $\bel=0$ and $\Psi^z_{\bar z}=0$.

Making use of the fact that $s(\bar\phi_{zz} \Psi^{\bar z}_{\bar z}) = \bar
\phi_{zz} \partial_z \Phi^z = 2\gamma^{\demi,0}\bar
\phi_{zz}\partial_z(\gamma^{\demi,0})$, the holomorphic part of the gravity
lagrangian (\ref{bel})  can be identified with the  $s$-exact expression
 $s (\bzz\bel+\bar \phi_{zz}\Psi^z_{\bar z})$   provided the superconformal
antighost is identified as
\begin{equation}\label{antighi}
\gz = 2\bar \phi_{zz}\gg.
\end{equation}
The field definitions  (\ref{redefs}) and (\ref{antighi}) involve
multiplication of fields by $\gg$   in a manner that implements the twists
needed to express the gravitational part of the action in a fully topological
form (as in \cite{twist}).  If the  twists on the fermionic matter fields are
implemented by an analogous change of field variables   the complete action
$I_B$ assumes the standard form of a topological $\sigma$  model coupled to
topological two-dimensional
gravity.

Before gauge fixing, the  topological theory does not contain specific
information about the target-space metric associated with any particular string
theory vacuum.  Of course, our gauge-fixed  derivation of the action $I_B$ is
background dependent  --   we chose  the background to be flat,  but it
presumably could have been more general.
 In a curved space-time endowed with  a closed 2-form
$\o=\o_{\m\n}dX^\m \wedge dX^\n$   the invariant
$I_{top}$ can be written as $I_{top}=\int d\o$. The
complexification of the coordinates involved in
(\ref{dec}) as well as  the relation between the NSR
fields and the topological ghosts and antighosts  can be obtained  by using
$\o_{\m\n}$ to define the polarizations.  Such a  generalisation of  the
construction to curved backgrounds should give a BRST-exact term that depends
covariantly on  the background metrics along the lines defined in
\cite{wittensigma}
\cite{bssigma}.      In these more general backgrounds the  distinction between
the r\^oles of type A and type B twistings in the type IIA and type IIB
superstring theories should become important.

The action (\ref{newprime}) is asymmetric with respect to the holomorphic and
anti-holomorphic sectors. To obtain a completely symmetrical formulation  a
further  fermionic field, $\rh'$,  could be introduced as a holomorphic partner
to
$\rh$. Together with its topological ghost, antighost and Lagrange multiplier
fields it  would have the decoupled action $I^{\prime\prime\prime} = \int d^2
z\sum_{\a=1}^{16} s(\Pb_{\rh'}^\a \pzb{\rh'}^\a)$ analogous to
(\ref{primedact}).
In the type II theory  the fields of the $\rh$ and $\rho'$ sectors decouple  so
that a different gauge choice could have been made where $\rh=\rh'= 0$.  This
is achieved in a BRST invariant
way by replacing $I^{\prime\prime}$   by $\int d^2 z\sum_{\a=1}^{16}
s(\Pb_\rh^\a \rh^\a)=
\int d^2 z\sum_{\a=1}^{16}  (
\l_\th^\a \rh_\rh^\a-\Pb_\rh^\a \P_\rh^\a)$ with a similar expression replacing
$I^{\prime\prime\prime}$.

\section{Relationship to space-time supersymmetry}

The GS superstring action for the type II  theories  involves the fields $X^\m$
and two space-time Weyl-Majorana fields   $\th^{Aa}$  ($A=1,2$, $a=1,\dots,16$)
that are world-sheet scalars \cite{GS}.   These two fields correspond to the
two space-time supersymmetry which have the same space-time chirality in the
type IIB theory and opposite chiralities in the type IIA theory.   In the
heterotic case there
is  one $\th^a$ field as well as the usual heterotic fermions, $\rho^\alpha$.
In
its original  derivation, the   GS   action    has  no  dependence on
the NSR fields.   It can
be   interpretated as a  nonlinear sigma model with a Wess--Zumino term
associated with super-Poincare invariance in the  target space \cite{henneaux}.

 In view of the results of last section, where both the type II and heterotic
models were related  to a  topological $\s$-model, it is tempting to identify
the fermionic  variables $\rh^\a$ as the 16 independent components of the
Majorana-Weyl spinor field  $\th^a$, up to a twist to accomodate  the change of
the
worldsheet   conformal weights from zero to one  half.
The doubling of the $\rh$ sector fields to accomodate
the existence of a pair of GS fields is an obvious possibility. In this way,
one
can imagine    promoting the following
infinitesimal topological transformations to the rank of a fundamental gauge
symmetry,
\be \label{fonde}
\d X^\m(z,\z)&=&\e^\m_X(z,\z)-i{\tilde\th^A}\g^\m\e^A_\th \nn\\
\d \th^{Aa}(z,\z)&=&\e^{Aa}_\th(z,\z).
  \ee
These  transformations extend (\ref{fond})  by taking into account  local
space-time supersymmetry transformations of  $X^\m$  in an equivariant way,

The associated BRST symmetry is
\be\label{fin}
sX^\m  &=& i\P^\m _X-{\tilde\th^A}\g^\m\P^A_\th
\nn\\
s\P^\m _X &=& 
{\tilde {\P}^A_{\th}}\g^\m \P^A_\th
\nn\\
s\Pb _X^\m  &=& \l_X^\m - {\tilde{ \Pb}^A_{\th}}\g^\m \P^A_\th
\nn\\
s\l^\m _X&=&  i{\tilde\l^A_{\th}}\g^\m \P^A_\th
\ee
and
\be
s\th^{A\a}  &=& i\P^{A\a}_\th
\nn\\
s\P^{A\a}_\th &=& 0
\nn\\
s\Pb^{Aa} _\th &=& \l^{Aa }_\th
\quad\quad
s\l^{Aa }_\th=0
\ee
(we have defined the Dirac conjugation ${\tilde\rh}=  \rh^\dagger \g^0$).

Although it is not central to  this paper it is noteworthy that  if one defines
the zero curvature
Cartan one-form,
\be
 (dX^\m-i{\tilde\rh}\g^\m d\rh)P_\m+d\rh^a Q_a,
\ee
where $P_\m$ and $Q_a$ are the generators of the N=1 super-Poincare
symmetry of  the target space, this BRST symmetry can be expressed as
\be
\left( (d+s)X^\m-i{\tilde\rh}\g^\m (d+s)\rh  \right )P_\m+(d+s)\rh^a Q_a=
i\P^\m_X P_\m+
i\P^a_\rh Q_a.
\ee
This equation  (and its Bianchi identy which determines the way $F^\mu_X$ and
$F^ a_\rho$ transform)   may be important in giving a geometrical
interpretation of
the topological ghosts, and understanding the
 meaning of  the topological term  $\int \o_{\m\n}dX^\m dX^\n$ in
10 dimensional space-time.

The form of the BRST transformations (\ref{fin}),  suggests that the
fundamental symmetry
of the theory   encodes general covariance and local
supersymmetry in the target space.
Heuristically, one  can think
of target-space as separated into all   possible 2-D surfaces. Thus, if one
builds  a theory based on the gauge symmetry  (\ref{fonde})  on each of these
surfaces,
and  then sums over them by integrating  over all classes of conformally
invariant parts of the two-dimensional metrics and gravitini,  one formally
reconstructs the  symmetry of N=2 supergravity,   $\d X^\m =\e^\m (X)
-i{\tilde\th^A}\g^\m\e^A_\th(X)$, $\delta \theta^A = \epsilon_\theta^A$.

However,  the problem of expressing the GS action as a combination of
BRST-exact and $d$-exact terms  remains open.    It should be possible to
formulate the theory as a topological model based on the above BRST symmetry in
such a way that the usual fermionic $\kappa$ symmetry emerges as a residual
local symmetry after gauge-fixing.   The $\kappa$ symmetry allows half of the
components of $\theta^{Aa}$ to be eliminated in passing to the light-cone gauge
by setting  $n_\mu \gamma^\mu \theta^A =0$, where $n^\mu$ is a null vector.  In
this respect it is intriguing that a null vector naturally arises in the BRST
system since  the variation,   $s \P_X^\mu =  {\tilde
{\P}^A_{\th}\g^\m \P^A_\th}$,  in (\ref{fin}) is a null vector (due to the
well-known properties of gamma matrices in $3,4,6$ and $10$ dimensions).

 The distinction  between the type II  and   heterotic  models in the GS
formulation  should reside in different gauge choices  for  the $N=2$
supercoordinates $\th^1$ and $\th^2$.      Thus, in the heterotic case the
fermions for the internal symmetry would be identified with one set of fields,
$\theta^1$, $\lambda_\theta^1$ with the same gauge function as in the last
section, while $\theta^2$ would be associated with $N=1$ space-time
supersymmetry.

\end{document}